\documentclass[11pt]{article}
\usepackage[utf8]{inputenc}
\usepackage{caption}
\usepackage{braket,slashed,bm}
\usepackage{jheppub}
\usepackage{simplewick}
\usepackage{array,multirow}
\usepackage{amsmath} \DeclareMathAlphabet{\mathpzc}{OT1}{pzc}{m}{it}
\usepackage[normalem]{ulem}
\usepackage{calligra}
\usepackage{floatrow}
\DeclareMathAlphabet{\mathpzc}{OT1}{pzc}{m}{it}
\usepackage{mathrsfs}
\usepackage{tikz}
\usepackage{verbatim}
\usepackage{slashed}

\usepackage{cleveref}
\crefname{section}{Sec.}{Secs.}
\crefname{table}{Tab.}{Tabs.}
\crefname{figure}{Fig.}{Figs.}
\crefname{equation}{Eq.}{Eqs.}
\crefname{appendix}{Appendix}{Appendix}

\usepackage{pgf}

\usepackage{subcaption}
\usepackage{lscape}
\usepackage{hyperref}
\usepackage{graphicx}
\usepackage{booktabs}
\newfloatcommand{capbtabbox}{table}[][\FBwidth]

\newcommand{\frules}{{\sc Feyn\-Rules }}

\newcommand{\ra}[1]{\renewcommand{\arraystretch}{#1}}  

\graphicspath{{./Figures/}}

\newcommand{\met}{\slashed{p}_T}
\def\beq{\begin{equation}}
\def\eeq{\end{equation}}
\def\bea{\begin{eqnarray}}
\def\eea{\end{eqnarray}}

\pgfkeys{/pgf/number format/.cd ,precision=2,sci generic={exponent={\times 10^{#1}}}}

\allowdisplaybreaks

\begin{document}
\preprint{KIAS - A25019}

\title{Hunting and identifying coloured resonances in four top events with machine learning}

\affiliation[a]{Center for AI and Natural Sciences, KIAS, Seoul 02455, Korea}

\affiliation[b]{Department of Physics, Chungbuk National University, Cheongju, Chungbuk 28644, Korea}

\affiliation[c]{Institut f\"{u}r Theoretische Physik und Astrophysik, Uni W\"{u}rzburg, Emil-Hilb-Weg 22, D-97074 W\"{u}rzburg, Germany}

\author[a]{Thomas Flacke,}
\emailAdd{flacke@kias.re.kr}

\author[b]{Jeong Han Kim,}
\emailAdd{jeonghan.kim@cbu.ac.kr}

\author[c]{Manuel Kunkel,}
\emailAdd{manuel.kunkel@uni-wuerzburg.de}

\author[b]{Jun Seung Pi,}
\emailAdd{junseung.pi@cbu.ac.kr}

\author[c]{Werner Porod}
\emailAdd{porod@physik.uni-wuerzburg.de}

\abstract{
We study prospects to search for pair or singly produced colour octet or colour sextet scalars which decay into two top quarks at the LHC. We focus on the same-sign lepton final state. We train a neural network comprising a simple multilayer perceptron combined with a convolutional neural network to optimize the separation of signal and background events. For LHC operated at 14 TeV and a luminosity of 3 ab$^{-1}$ we find an expected discovery reach of $m_8=1.8$~TeV and $m_6=1.92$~TeV for pair produced colour octets and sextets, respectively, and an expected exclusion reach of $m_8=2.02$~TeV and $m_6=2.14$~TeV.
In a second step, we retrain the same network architecture to discriminate between signal processes.
The network can clearly distinguish between the different colour representations. Moreover, we can also determine whether there is a significant contribution from single production to pair production for the same final state.
The methodology can be applied to BSM candidates of different spin and colour representations.}
\maketitle


\section{Introduction}
\label{sec:intro}

The Standard Model (SM) of particle physics has turned out to be a very successful description of Nature up to the TeV scale. 
However, it leaves several problems open: (i) What causes the hierarchies in the fermion sector, both in view of their masses and mixing patterns? (ii) What is the nature of the observed dark matter? (iii) What stabilises the Higgs potential against radiative  corrections? 
These questions have motivated the exploration of models beyond the SM (BSM). Many of these extensions predict the existence of new states in a variety of charge and colour representations.
New states will have large production cross sections at the LHC up to the TeV range if they are colour charged. Typical examples are squarks and gluinos in supersymmetric models,
leptoquarks, or top-partners in Composite Higgs models (CHM). 

Provided a new state is discovered at the LHC, it will be ultimately important to determine its properties, such as mass, spin, parity, and gauge quantum numbers. 
The most challenging of all is the determination of the colour quantum number as it cannot be measured directly. 
Coloured BSM states are predicted by various
extensions of the SM:
(i) In supersymmetric models every SM quark has scalar partner with the same quantum numbers assigned, and the gluons have fermionic partners \cite{Martin:1997ns}.  In N=2 supersymmetry inspired models the latter have in addition scalar colour octet partners, see e.g.~\cite{Benakli:2014cia} and refs.~therein.
(ii) Composite Higgs models with partial compositeness \cite{Kaplan:1991dc} require top-partners to explain the largeness of the top-mass compared to the other quark masses. In models with a fermionic UV completion \cite{Ferretti:2013kya,Ferretti:2016upr} one can also have fermionic resonances
 in the colour sextet and octet representation, see e.g.~\cite{Cacciapaglia:2021uqh}. These models predict in all cases coloured pseudo Nambu-Goldstone bosons in the octet representation, but additional sextets or triplets are also possible \cite{Cacciapaglia:2015eqa,Ferretti:2016upr,Krause:2018cwe,Cacciapaglia:2019bqz,Cacciapaglia:2020kgq}. Moreover, they also predict coloured spin-1 resonances  coming as colour triplets, sextets and octets \cite{Cacciapaglia:2024wdn}.
(iii) Lepto-quark models predict colour triplet scalars, see \cite{Dorsner:2016wpm} and refs.~therein,  but UV completions can also contain scalar octets, see e.g.~\cite{FileviezPerez:2013zmv,Faber:2018qon,Faber:2018afz,FileviezPerez:2022rbk}. %
(iv) Models with extra space dimensions predict Kaluza-Klein exitations of the SM particles, e.g.~KK gluons (spin-1 colour octets) \cite{Davoudiasl:1999tf,Pomarol:1999ad,Chang:1999nh,Bajc:1999mh,Cheng:2002ab,Lillie:2007yh}.

Particles with different colour structures give rise to distinct colour flows, leading to unique radiation patterns that can be used to construct observables for differentiation. The classification of colour octet and singlet dijet resonances has been studied in refs.~\cite{Ellis:1996eu,Gallicchio:2010sw,Atre:2013mja} and the distinction between colour triplet and colour sextet dijets in \cite{Chivukula:2015zma}.
The initial and final state quarks are colour connected in case of a colour octet resonance, leading to increased QCD radiation in the forward direction. In contrast, for a singlet resonance, there is no correlation with the initial state beams; instead, the final state quarks are colour-connected with each other, resulting in increased radiation along the direction of the quark-antiquark pair. 
Reference \cite{Han:2023djl} provided an in-depth study on distinguishing coloured dijet resonances using machine learning techniques. Similar methods have been employed in the search for doubly charged pNBGs \cite{Flacke:2023eil}, double Higgs production \cite{Kim:2019wns, Huang:2022rne}, and top tagging, see  \cite{Kasieczka:2019dbj} and refs.~therein. Deep neural networks have also been explored for distinguishing between colour sextet and octet states \cite{Chakraborty:2019imr}. Additionally, experimental analyses have utilised colour flow to distinguish different processes, such as $t \bar{t}$ measurements at the Tevatron \cite{D0:2011lzz} and by ATLAS \cite{ATLAS:2015ytt, ATLAS:2018olo, ATLAS:2023jdw}.

In this paper, we focus on the production of scalars coming either as colour octet or as colour sextet. We assume that they dominantly decay into top quarks.
They can be either singly or pair-produced, leading in all cases to final states with four top quarks, albeit with different kinematics and/or colour flow. 
We will demonstrate on the one hand that these particles can be detected at the high luminosity LHC (HL-LHC) even for masses close to 2 TeV. 
On the other hand, we will show that we can distinguish the colour representation even for a mass of 1.8~TeV which is close to the reach of HL-LHC. 
The paper is organized as follows: we discuss the signal processes in the next section and in \cref{sec:eventgen} the background processes and the preselection cuts. 
In \cref{sec:input} we present the data representation and the network architecture of our ML approach
and give the corresponding results in \cref{sec:results}. In \cref{sec:conclusion} we draw our conclusions and give an outlook. Additional details are given in  \cref{app:single,app:sep,sec:appen}.

\section{Signal processes} 
\label{sec:model}

In minimal CHMs with an underlying fermionic description \cite{Ferretti:2013kya,Ferretti:2016upr,Belyaev:2016ftv}, an electrically neutral colour octet $S_8$ emerges as a generic prediction.
Half of the twelve promising models identified in \cite{Belyaev:2016ftv} further contain a colour sextet scalar $S_6$ with charge $4/3$.
In the other models, the $S_8$ can be accompanied by a stop-like colour triplet with charge $2/3$ or a sextet with charge $-2/3$.
We do not study the latter in this work because only the former can facilitate the production of top quarks: 
Through partial compositeness interactions the $S_8$ and $S_6$ obtain couplings to $t\bar t$ and $tt$, respectively.
We parameterize them as follows using a simplified model ansatz:
\begin{align}
    \mathcal L = &\frac 12 (D_\mu S_8)^T D^\mu S_8 - \frac 12 m_8^2 \, S_8^a S_8^a + \lambda_8 S_8^a \, \bar t \, (i\gamma_5) t^a_{\mathbf 3} \, t \label{eq:models8} \\
    &+ (D_\mu S_6)^\dagger D^\mu S_6 - m_6^2 \, S_6^\dagger S_6 + \lambda_6 \left( K^s_{ij}\,  S_6^s\, \bar t_i P_L t_j^c + \mathrm{h.c.} \right) \label{eq:model6}
\end{align}
Here $t_{\mathbf 3}^a = \frac 12 \lambda^a$ are the SU(3) generators in the fundamental representation and $K^s_{ij}$ are the Clebsch-Gordon coefficients connecting a sextet to two triplets.
The $S_8$ is a pseudo scalar, and the $S_6$ couples to right-handed top quarks. 
In the following we will set $m_8=m_6\equiv m_S$ for simplicity.

\begin{figure}[]
    \centering
    \includegraphics[width=0.31\linewidth]{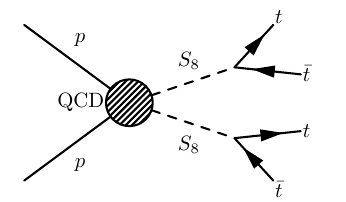}
    \includegraphics[width=0.31\linewidth]{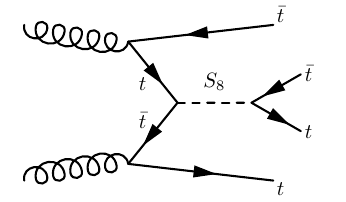}
    \includegraphics[width=0.31\linewidth]{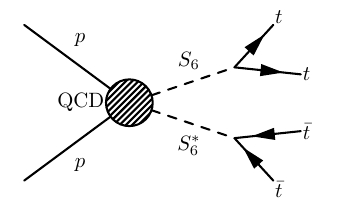}
    \caption{Production of four top quarks via the QCD pair productions and the single production of coloured scalars. }
    \label{fig:feynman_4t_v2}
\end{figure}

In models with partial compositeness, the decays into top quarks typically dominate as these couplings are proportional to the corresponding quark mass.
Both states have subleading decays in CHMs: $S_8 \to gg,\,gZ,\,g\gamma$ through the Wess-Zumino-Witten terms, and partial compositeness-induced decays into lighter quarks for both $S_8$ and $S_6$.
However, in the following we assume $\mathrm{Br}(S_8 \to t\bar t) = 1 = \mathrm{Br}(S_6 \to tt)$ for simplicity.
Both states then yield four top quarks when pair or singly produced:
\begin{align}
    &pp \to S_8 S_8 \to t\bar tt\bar t \label{eq:S8S8}\\
    &pp \to S_8 t\bar t \to t\bar tt\bar t \label{eq:S8}\\
    &pp \to S_6 S_6^* \to tt \bar t\bar t \label{eq:S6S6}
\end{align}
These are the signal processes we target in this work. 
We show the corresponding Feynman diagrams in \cref{fig:feynman_4t_v2}.
The $S_6$ can be produced singly analogously to the octet, but we do not study sextet single production in this work as we do not expect to gain new insights from it compared to the octet.

\begin{figure}
    \centering
    \includegraphics[width=0.55\linewidth]{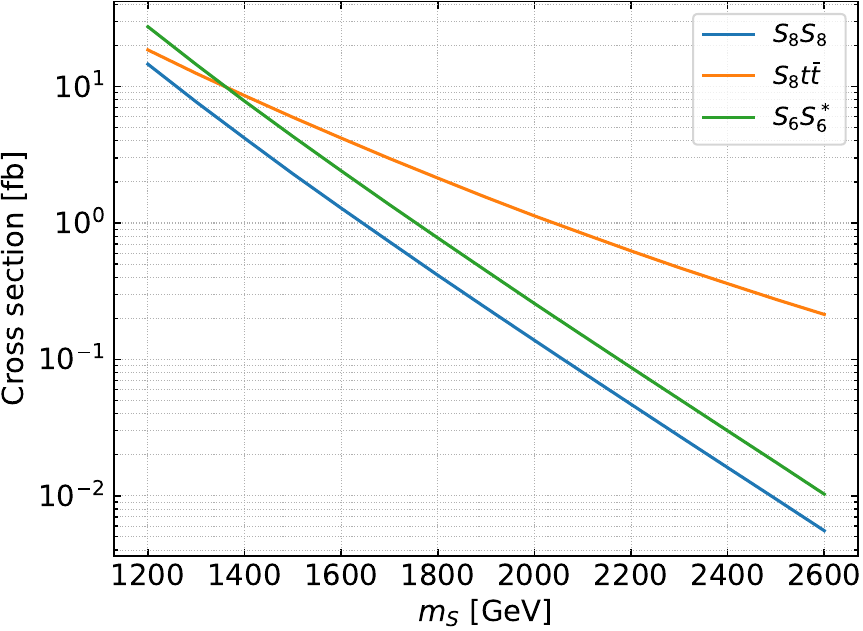}
    \caption{Leading order cross sections for producing four top quarks via coloured scalars at $\sqrt{s} = 14$~TeV.
    We show the cross sections for sextet and octet pair production and octet mixed production with coupling $\lambda_8 = 1.1$ for which the total width is slightly less than 10\% of the mass.  }\label{fig:xs1dim}
\end{figure}

We generated UFO libraries \cite{Degrande:2011ua} from the public \frules \cite{Alloul:2013bka} implementations of octet \cite{fukssgluonufo,Darme:2018dvz,Darme:2021gtt} and sextet scalars \cite{alwalldiquarkufo,Han:2009ya} to calculate the cross sections of the signal processes. 
We use {\sc MadGraph5\_aMC@NLO} \cite{Alwall:2014hca,Hirschi:2015iia} v3.5.3 with with NNPDF2.3QED parton distribution functions \cite{Ball:2013hta} at a centre-of-mass energy of $\sqrt s=14$~TeV to calculate leading order (LO) cross sections.
The renormalisation and factorisation scales were set to $\mu_R = \mu_F = m_S$. 
Given that no NLO-capable UFO is available for the sextets, we conservatively work with leading order cross sections for both sextet and octet production to allow for better comparison. 
In \cref{fig:xs1dim} we show the resulting cross sections cross sections of the signal processes.
The lines for the pair production of $S_6 S^*_6$ and $S_8 S_8$ correspond to the first two diagrams shown in \cref{fig:feynman_4t_v2}. In case of single $S_8$ production there is, however, an ambiguity as diagrams with an off-shell $S_8$ contribute as well. If the partonic center of mass is sufficiently high, the second $S_8$ can become on-shell. Consequently, we calculate the sum of both single and pair production and choose a sufficiently large $\lambda_8$ such that the single production component clearly dominates over the pair production. In the following, we will take $\lambda_8=1.1$ as our particular choice. On the one hand, it allows for the desired dominance. On the other hand, the width of the $S_8$ is just below 10 per-cent of its mass which allows us to use still a narrow width approximation\footnote{Broad resonances require a different analysis, see e.g.~\cite{Jung:2019iii} for resonances decaying to $t\bar{t}$ and~\cite{Deandrea:2021vje} for vector-like top-partners.}. We provide more details on this aspect in \cref{app:single}.

\section{Preselection cuts and backgrounds}
\label{sec:eventgen}

We study three signal processes, each leading to four top quarks.
We separate them from their Standard Model backgrounds in a two step approach:
we first employ standard kinematic cuts to remove a large portion of the backgrounds.
The events that pass this preselection are then used for a more sophisticated deep learning analysis.

A very effective approach to reducing the SM backgrounds is to target the final state with two same-sign leptons,
\begin{align}
    t\bar tt\bar t \to (W^+ b) (W^-\bar b) (W^+ b) (W^- \bar b) \to \ell^\pm \ell^\pm + 4b+4j+\slashed{p}_T,
\end{align}
since this eliminates the significant $t\bar t+$jets backgrounds.
To pass preselection we demand events to satisfy the following conditions:
\begin{itemize}
    \item exactly two same-sign isolated leptons with possibly different flavour, each with $|\eta|<2.5$ and $p_T > 20$~GeV
    \item each lepton must satisfy the isolation condition $p_T(\ell)/(p_T(\ell) + \sum_i p_{T,i})>0.7$ where $\sum_i p_{T,i}$ represents the total sum of the transverse momenta of surrounding particles with $p_{T,i} > 0.5$~GeV and separated by $\Delta R_{i\ell} < 0.3$
    \item at least three light jets with $|\eta|<2.5$ and $p_T > 25$~GeV, reconstructed with anti-$k_T$ with cone radius $r=0.4$
    \item at least three b-tagged jets with $|\eta|<2.5$ and $p_T > 25$~GeV, reconstructed with anti-$k_T$ with cone radius $r=0.4$ 
    \item missing transverse momentum $\slashed{p}_T>20$~GeV
    \item $S_T > $ 400 GeV where $S_T$ is the scalar sum of the transverse momenta of the reconstructed jets and the two same-sign leptons
\end{itemize}
These match the cuts used in our previous work \cite{Flacke:2023eil}.

In light of these cuts the most important backgrounds are
\begin{align}
    t\bar t t\bar t, \quad t\bar th, \quad t\bar tV, \quad t\bar t VV,
\end{align}
where $V = W^\pm, Z$.
We generate hard scattering events for these backgrounds with {\sc MadGraph5\_aMC@NLO} at a centre-of-mass energy of $\sqrt{s} = 14$ TeV using NNPDF2.3QED parton distribution functions. 
We run the event generation at NLO accuracy in QCD apart from $t\bar tt\bar t$, for which we generate leading order events.
We obtain the NLO cross section values of $1.17 \times 10^{-2}$~pb ($t\bar tt\bar t$), $5.60 \times 10^{-1}$~pb ($t\bar th$) and $1.60$~pb ($t\bar tV$), and $1.89 \times 10^{-2}$~pb ($t\bar tVV$).
For the signal processes \cref{eq:S8S8,eq:S8,eq:S6S6}, we generate leading order events with the simulation setup described in the previous section.
The hard scattering events are then passed to {\sc Pythia8} \cite{Sjostrand:2014zea} for parton showering and hadronisation.

We now apply the preselection cuts to the hadronised signal and background events.
To this end, we simulate the detector response with {\sc Delphes} 3.4.1 \cite{deFavereau:2013fsa} based on modified ATLAS settings \cite{Kim:2019wns}. 
For jet reconstruction, we employ the anti-$k_T$ clustering algorithm implemented in {\sc Fastjet} 3.3.1 \cite{Cacciari:2011ma}.
The $b$-jets are tagged using a constant efficiency of $\epsilon_{b \rightarrow b} = 0.8$. The misidentification rates for $c$-jets and light-flavor jets being tagged as $b$-jets are set to $\epsilon_{c \rightarrow b} = 0.2$ and $\epsilon_{j \rightarrow b} = 0.01$, respectively \cite{CERN-LHCC-2017-021}.
We do not take into account lepton charge misidentification as it is at most a percent level correction \cite{CERN-LHCC-2017-021}.
In \cref{tab:Basiccut} we list the resulting efficiencies, cross sections, and the number of events assuming the HL-LHC dataset of 3~ab$^{-1}$.
These events and cross sections are the basis for the remainder of this work.

\begin{table}[]
	\ra{1.15}
	\begin{tabular}{lccc}
		\toprule
		Process & $\epsilon_\mathrm{Preselection}$ & Cross section [fb] & Events at 3~ab$^{-1}$ \\ \midrule
		$S_8 S_8$ & $1.71\times 10^{-2}$ & $3.91\times 10^{-2}$ & 117\\
		$S_6 S_6^*$ & $1.71\times 10^{-2}$ & $7.37\times 10^{-2}$ & 221 \\
        $S_8 t \bar t$ & $1.83\times 10^{-2}$ & $1.51\times 10^{-1}$ & 452 \\ \midrule
		$t\bar t V$        & $1.70\times 10^{-4}$             & $2.72\times 10^{-1}$ & 816                  \\
		$t\bar th$         & $3.75\times 10^{-4}$             & $2.10\times 10^{-1}$ & 629                  \\
		$t\bar tt\bar t$   & $1.63 \times 10^{-2}$            & $1.91\times 10^{-1}$ & 572                  \\
		$t\bar t VV$       & $1.74\times 10^{-3}$             & $3.29\times 10^{-2}$ & 98                   \\
		$VVV$              & $2.08\times 10^{-6}$             & $1.05\times 10^{-3}$ & 3                    \\ \bottomrule
	\end{tabular}
	\caption{Signal and background efficiencies and cross sections after the preselection cuts. For the signal processes we take a reference scalar mass of $m_S = 1.5$ TeV and $\lambda_8=1.1$.}
	\label{tab:Basiccut}
\end{table}

\section{Data representation and network architectures}
\label{sec:input}
The next task is to find a useful representation of the data and network architectures to efficiently separate the signal from the background after selecting the events. 
In ref.~\cite{Flacke:2023eil} we have shown that for a similar signature, a combination of jet images constructed from low-level calorimeter data and high-level kinematic variables of the reconstructed objects allows for an excellent discrimination between signal and background. We demonstrate in the following that this also holds for the case at hand.

We build a first data set $\mathcal K$ from kinematic variables constructed from the two same-sign leptons, the three leading $b$-tagged jets, and the three leading non-$b$-tagged jets. 
The kinematic variables are represented by 
\begin{equation}
\mathcal K = \bigcup_{i \neq j} M_{ij} \cup \bigcup_{i \neq j }\Delta R_{ij} 
\cup \bigcup_{i } p_{Ti} \cup \{ \met, S_T \} \;  ,\label{eq:5kin}
\end{equation}
where $M_{ij}$ and $\Delta R_{ij}$ denote the invariant mass and angular separation between two reconstructed objects $i$ and $j$, $p_{Ti}$ represents the transverse momentum of the object $i$, $\met$ is the missing transverse momentum, and $S_T$ is the scalar sum of the transverse momenta of the reconstructed jets and the two same-sign leptons.
No normalization is applied to the kinematic data.

In addition we construct a second data set consisting of jet images, which are constructed from the particle flow data from each event and are acquired through {\sc Delphes}, following a procedure similar to that described in Ref. \cite{Huang:2022rne, Flacke:2023eil, Kim:2019wns}. 
We begin by separating particle flow information into charged and neutral components. 
Charged particles primarily consist of charged hadrons, whereas neutral particles include neutral hadrons and non-isolated photons. 
Importantly, isolated leptons are excluded from both groups.

We now project the particles into the $\eta$-$\phi$-plane, with its origin defined at the geometric center of the two isolated same-sign leptons in the event. 
To preserve relevant spatial features, periodicity is enforced in both the $\eta$ and $\phi$ directions. 
Although $\phi$ naturally exhibits periodicity, $\eta$ does not. 
However, we treat $\eta$ as artificially periodic to avoid either the loss of data or the need to expand images with largely empty regions.
The resulting jet images are divided into a grid of calorimeter cells spanning the region $-2.5 \leq \eta \leq 2.5$ and $-\pi \leq \phi \leq \pi$. 
Each pixel’s intensity reflects the total transverse momentum of all particles that fall within that specific pixel. Only particles with a transverse momentum greater than 0.7 GeV are considered. 
We choose not to normalize the pixel intensity values, as the actual magnitude of the transverse momentum can enhance the differentiation among signals and backgrounds.
The jet image data is organized into a $(2 \times 50 \times 50)$ data structure, with one channel each representing charged and neutral particles. 
Additionally, to incorporate spatial correlations among the final state particles, we project the two isolated same-sign leptons onto the $50 \times 50$ grid, integrating their information into the jet images. 
As a result, our final dataset has dimensions of $\mathcal I_{CN\ell} = (3 \times 50 \times 50)$. 

A significant challenge in utilizing particle flow data arises from pileup contamination, especially in the high pileup conditions anticipated at the HL-LHC, with an average of $\langle \mu \rangle \sim 200$ pileup interactions per bunch crossing \cite{ATLAS:2019xli}. 
For an in-depth analysis of pileup dynamics and various mitigation techniques, we refer readers to Ref. \cite{Kim:2019wns}, which presents a semi-realistic assessment of these challenges. 
Although longitudinal vertex information can substantially reduce the contamination of charged particles from pileup \cite{Bertolini:2014bba}, this approach is not effective for neutral particles. 
Nevertheless, experimental efforts are actively being made to mitigate the effects of neutral pileup contamination \cite{CMS:2020ebo,Bertolini:2014bba}. 
In our previous work \cite{Flacke:2023eil} we accounted for this by performing a separate analysis excluding the neutral jet image, which yielded a discovery reach that was weaker than from the full image data but stronger than using only kinematic data. We expect the same to hold in the present case. 
Furthermore, we have reason to expect this difference to be smaller in the present work, which we will discuss in \cref{sec:results}.

\begin{figure}
    \centering
    \includegraphics[width=0.9\linewidth]{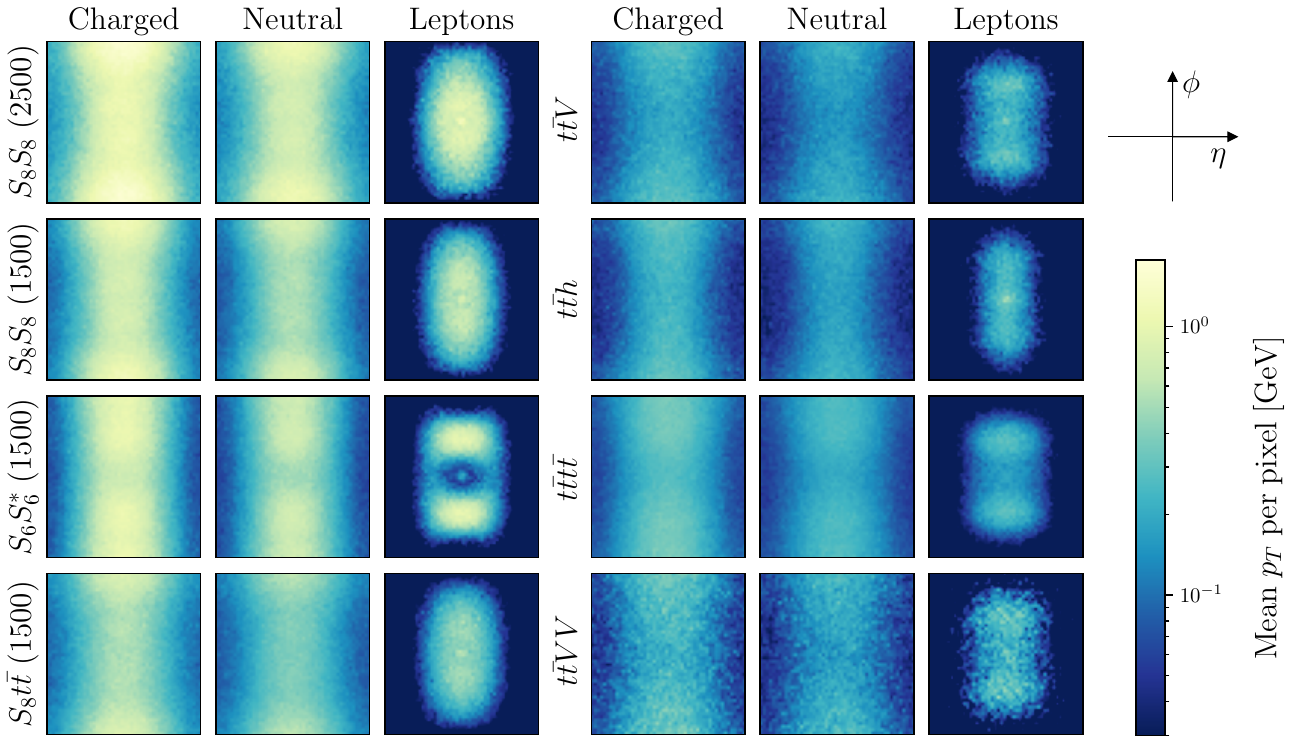}
    \caption{Jet images for background (right) and signal processes (left) where the parentheses indicate the scalar mass in GeV. The three columns show the images of charged hadrons, neutral hadrons, and the two isolated leptons. Each panel shows the average distribution of particles taken over all simulated events.}
    \label{fig:jetimages}
\end{figure}

In \cref{fig:jetimages} we present the jet images compiled from all simulated events, where the left three columns represent the signal samples with the reference mass in GeV indicated in parentheses and the right three columns depict the background samples.
Comparing the signal and background jet images reveals that signal images are typically brighter, which indicates a larger total $p_T$. 
This brightness further increases as the scalar mass becomes larger.
In addition, for $S_6 S^*_6$ signal events, the leptons are generally concentrated near $\phi=\pm  \pi/2$, whereas they are more widely dispersed in the $S_8 S_8$ or $S_8 t\bar{t}$ events. This difference arises because, for the $S_8 S_8$ or $S_8 t\bar{t}$, the same-sign leptons are produced from the decays of different particles, resulting in greater angular separation. In contrast, in signal events for $S_6 S^*_6$ production, both same-sign leptons originate from either $S_6$ or $S^*_6$, leading to a more clustered distribution.

\begin{figure}
	\centering
    \includegraphics[width=0.7\textwidth,clip]{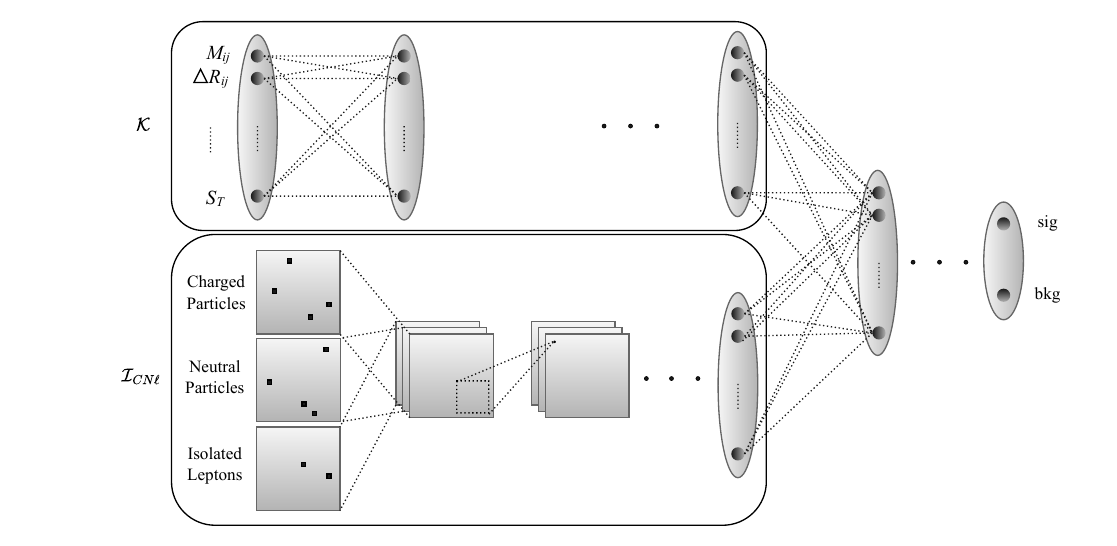}
    \caption{Schematic architecture of the neural network used in this article.}
    \label{fig:NN_CNLK} 
\end{figure}

Based on the jet images and kinematic variables, we employ neural networks to separate the signal from the SM backgrounds. Additionally, we extend our method to differentiate between the production channels of sextet and octet scalars, enabling a more precise identification of these distinct processes.
To achieve this, we utilize two neural network architectures.
The first is a simple multilayer perceptron (MLP) using kinematic variables $\mathcal K$.
As a more advanced architecture, we employ a combination of an MLP on $\mathcal K$ with a convolutional neural network (CNN) operating on jet images.
The schematic network structure is shown in \cref{fig:NN_CNLK}.
The two separate chains are interfaced to produce a combined output.
This combination proved to be more proficient than a pure CNN in ref.~\cite{Flacke:2023eil}, but for brevity, we will refer to it as just ``CNN'' in the following.
The full network architecture is detailed in \cref{sec:appen}.

Our data set is split into three parts: the training set (80,000 events) and the validation set (20,000 events) are comprised to equal parts of signal and background events with the latter processes appearing weighted by their cross section after the preselection cuts, while the holdout test set (80,000 events) has a higher proportion of signal events to allow for sufficient statistics for separating the signal processes.
We use the training set to optimize the network parameters using the cross entropy loss function and the {\sc Adam} optimizer. 
During training, we monitor the generalization error using the validation set and use the network configuration with the lowest validation loss for further analysis.
The results presented in the next section were obtained from the test set that was only used for final evaluation.
For each process, network, and mass point we trained 20 copies of the network to account for the systematic uncertainties associated with the limited number of events and the stochastic network training.
The lines in the following indicate the average results while the shaded regions indicate the $1\sigma$ error which is estimated from the variance over the 20 training runs.

\section{Results}
\label{sec:results}

We first apply our trained networks to test set events to obtain the discovery reach and exclusion limit at the HL-LHC.
In a second step, we show how our networks can be used to identify which resonance causes the excess if one is discovered. 

\subsection{Discovering a signal}

\begin{figure}
    \centering
    \begin{subfigure}{0.31\linewidth}
        \includegraphics[width=\linewidth]{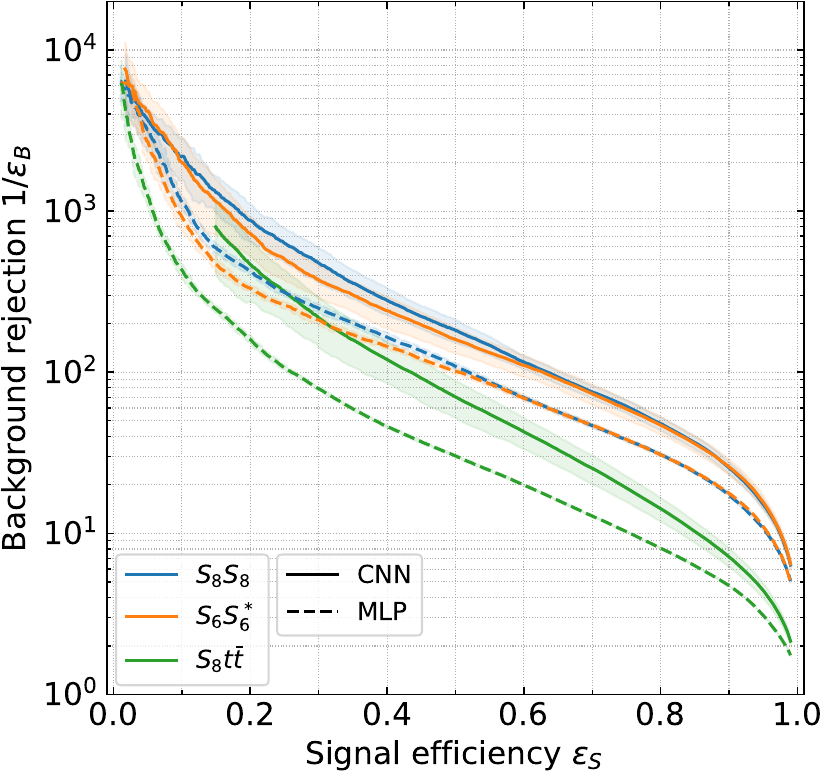}
        \caption{$m_S = 1200$~GeV}
    \end{subfigure}
    \begin{subfigure}{0.31\linewidth}
        \includegraphics[width=\linewidth]{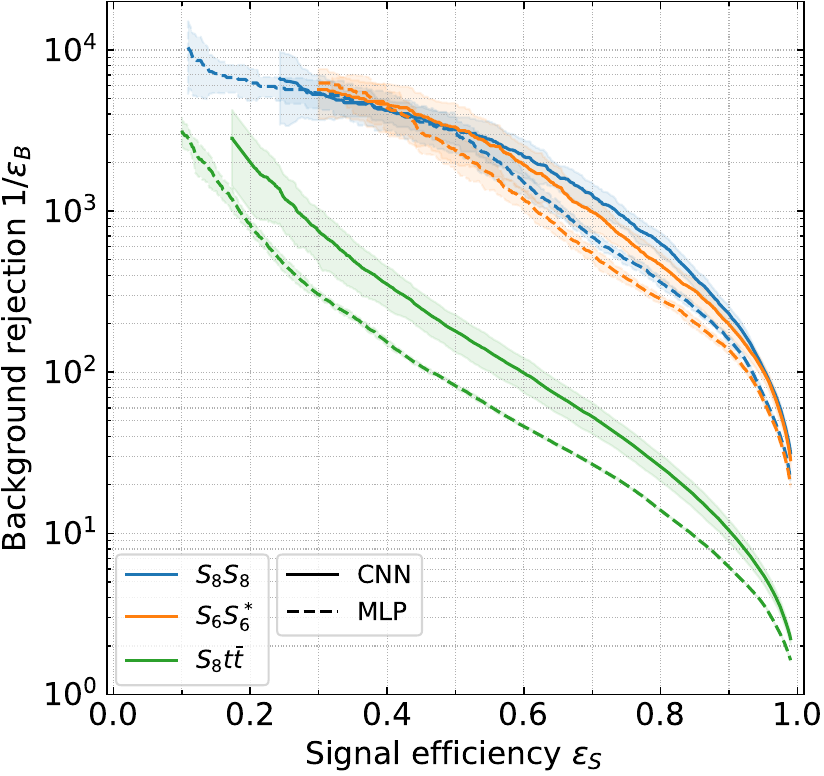}
        \caption{$m_S = 1800$~GeV}
    \end{subfigure}
    \begin{subfigure}{0.31\linewidth}
        \includegraphics[width=\linewidth]{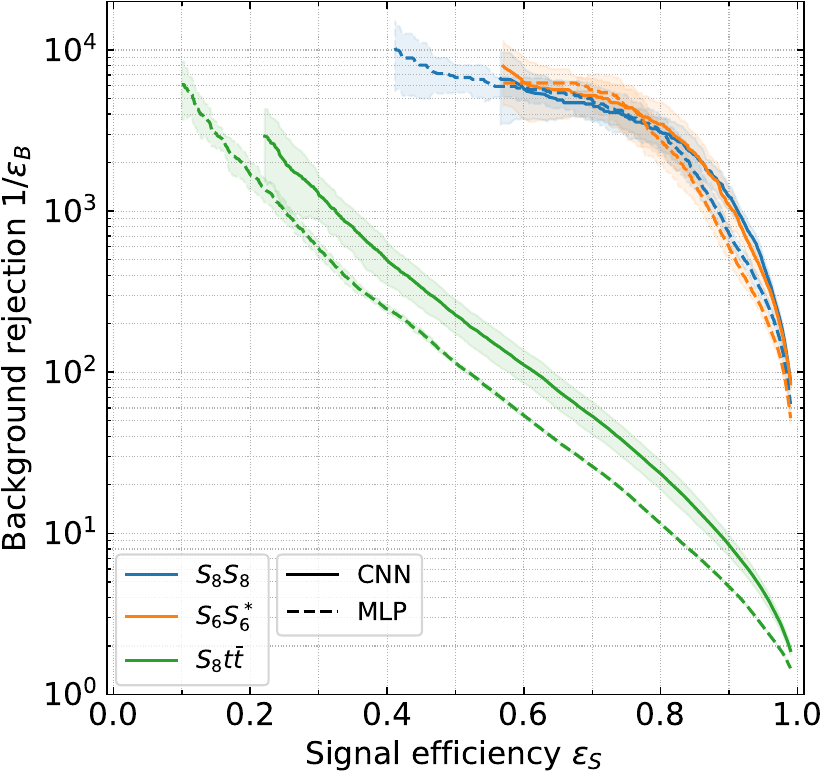}
        \caption{$m_S = 2300$~GeV}
    \end{subfigure}
    \caption{Receiver operator characteristic (ROC) curves comparing network performances for different processes and mass points.}
    \label{fig:roc}
\end{figure}

We show the receiver operator characteristic (ROC) curves in \cref{fig:roc} to assess the performance of our networks in separating signal from background events.
The ROC curves indicate how many background events can be successfully rejected for a given signal efficiency. 
We see that the discrimination is comparable for $S_8 S_8$ and $S_6 S_6^*$, while single production is distinguished significantly worse.
The discrimination increases with the scalar masses, which can be explained with the increased $p_T$ of the final state objects.
For all processes the MLP performs worse than the CNN at low masses but then catches up with increasing $m_S$.
At high masses both networks perform comparably for pair production while the CNN still provides a stronger discrimination for $S_8t\bar t$. 

We now complete our search design by placing a cut on the neural network (NN) score. 
One could determine this dynamically, e.g.\ such as to maximise a significance, but this tends to lead to a very small number of expected background events in our case.
We instead opt for a fixed NN score cut, demanding 5 background events. 
Passing this cut are $S$ signal events and $B=5$ background events.
From these we calculate the significances for discovery and exclusion by
\begin{equation}
Z_\mathrm{dis} \equiv
\sqrt{-2\,\ln\bigg(\frac{L(B | S\!+\!B)}{L( S\!+\!B| S\!+\!B)}\bigg)}, \qquad Z_\mathrm{exc} \equiv
\sqrt{-2\,\ln\bigg(\frac{L(S\!+\!B | B )}{L( B | B)}\bigg)}
\label{Eq:SigExc} 
\end{equation}
with the Poisson likelihood $L(x |n) =  \frac{x^{n}}{n !} e^{-x}$~\cite{Cowan:2010js}.
We rescale the signal cross sections until we reach
\begin{equation}
Z_\mathrm{dis} \geq 5, \qquad Z_\mathrm{exc} \geq 1.64 
\label{Eq:SigDis} 
\end{equation}
for discovery and exclusion, respectively, yielding an expected discovery reach $\sigma_{5\sigma}$ and an expected upper limit $\sigma_{95}$.

\begin{figure}
    \centering
    \begin{subfigure}{0.48\linewidth}
        \includegraphics[width=\linewidth]{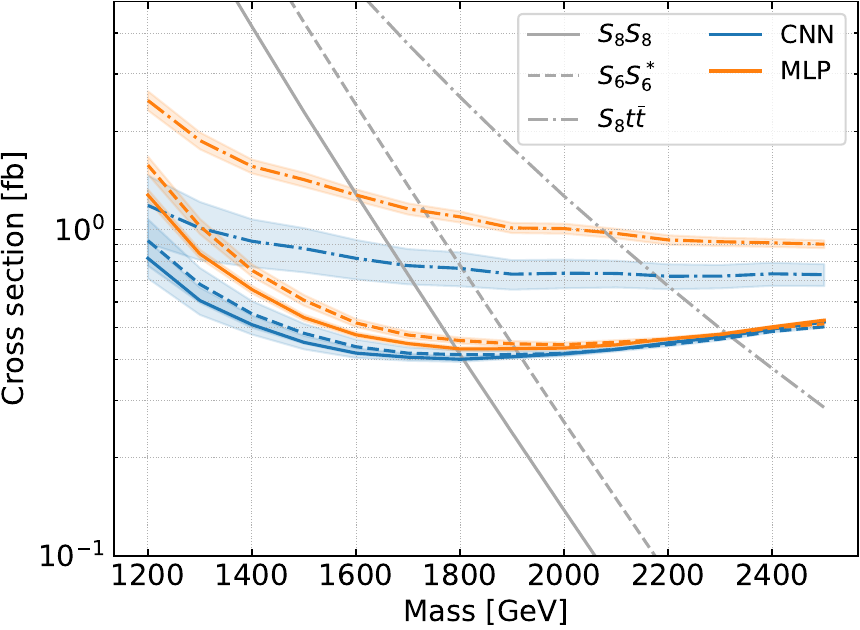}
    \end{subfigure}
    \begin{subfigure}{0.48\linewidth}
        \includegraphics[width=\linewidth]{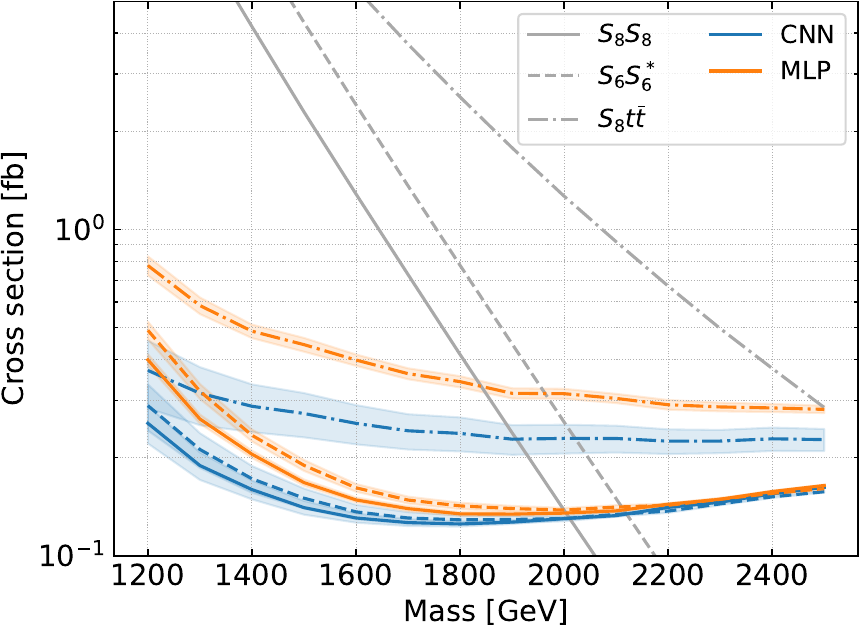}
    \end{subfigure}
    \caption{Expected discovery reach (left) and exclusion limit (right) for different processes and networks. The process $pp\to tt S_8$ is dominated by single production but also includes a portion of pair production.}
    \label{fig:results}
\end{figure}

The resulting limits for an assumed luminosity of 3~ab$^{-1}$ at LHC with $\sqrt{s}= 14$~TeV are shown in \cref{fig:results}. 
As can be seen, colour octets and sextets can be discovered for $m_8\leq 1.80$~TeV and $m_6\leq 1.92$~TeV based solely on pair production while, in absence of a signal, they can be excluded for $m_8\leq 2.00$~TeV and $m_6\leq 2.12$~TeV. Single production, which depends on the coupling $\lambda_8$, enhances the signal cross section, thus extending sensitivity to even higher masses, as shown for the process $pp\rightarrow S_8 t\bar t$ with $\lambda_8 = 1.1$ in \cref{fig:results} as an example.
As expected from the ROC curves in \cref{fig:roc}, the MLP and the CNN lead to comparable limits for pair production\footnote{We note that this implies that a separate analysis that excludes the pileup-ridden neutral jet images would not have yielded new information for pair production.} while the CNN outperforms the MLP for $S_8 t\bar t$.

The current exclusion limits on the masses of octet and sextet are $m_8\leq 1.38$~TeV and $m_6\leq 1.51$~TeV \cite{Kunkel:2025qld}, respectively, as determined from recasts of ATLAS SUSY searches \cite{ATLAS:2021fbt,ATLAS:2022ihe}.
Notably, a recent study of colour octet pair and single production \cite{Darme:2024epi} reached the same conclusion and almost identical projected exclusion reach ($m_8\leq 2.04$~TeV). Ref.~\cite{Darme:2024epi} studies not only the same-sign lepton channel but also includes two signal regions with a single lepton.  They do not make use of machine learning techniques and instead use highly boosted top tagging techniques. In particular, \cite{Darme:2024epi} use more advanced lepton isolation criteria than the present work which avoid excessive signal loss for very highly boosted top decays in which the leptons tend to be in proximity of the $b$ quark. The lepton isolation criteria used in the study presented here are more basic and not specifically tailored to highly boosted top quarks, which results in notable signal loss when applying our preselection criteria. The fact that we find an almost identical projected exclusion reach despite the lower preselection efficiency demonstrates the selection power of the neural network. It also indicates that combining boosted top adapted lepton isolation, boosted top tagging and a machine learning architecture as presented here is likely to allow for further improvement in sensitivity.

\subsection{Identifying the signal process}\label{sec:seperation}

We have shown so far that our neural networks are very proficient at separating the signal events from the SM backgrounds and have determined up to which masses the signal could be discovered with our method.
We now go one step further and assume that an excess in four top quark production which can be interpreted as one of our signal processes will be discovered at the HL-LHC.
There is significant cross acceptance among $S_8 S_8$, $S_6 S_6^*$, and $S_8 t\bar{t}$ (see \cref{app:sep} for more details). 
This raises the question if we can successfully tell the signal processes apart with our networks.
In this section we illustrate how this can be done using a scalar mass of 1.8~TeV as benchmark.

\begin{figure}
    \centering
    \begin{subfigure}{0.32\linewidth}
        \includegraphics[width=\linewidth]{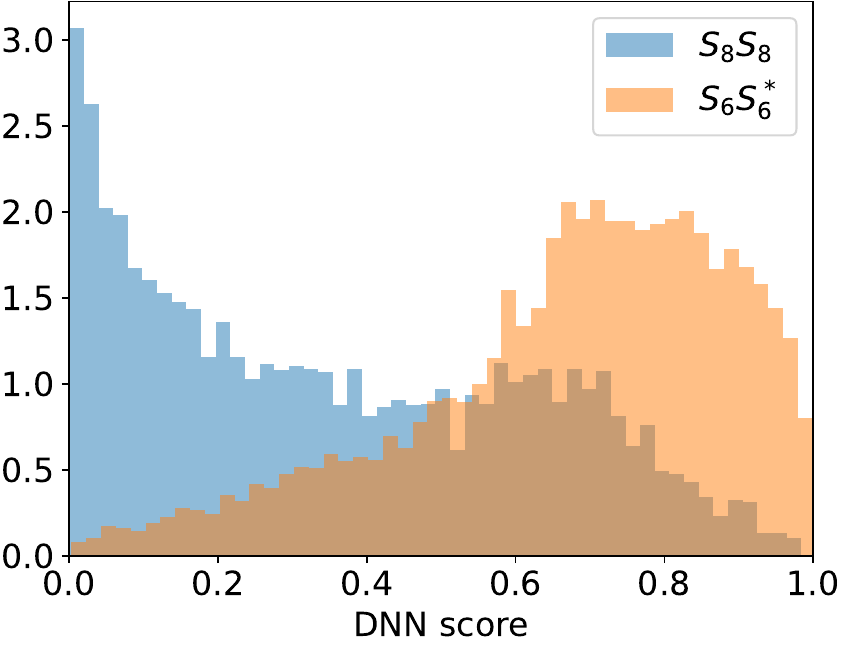}
    \end{subfigure}
    \begin{subfigure}{0.32\linewidth}
        \includegraphics[width=\linewidth]{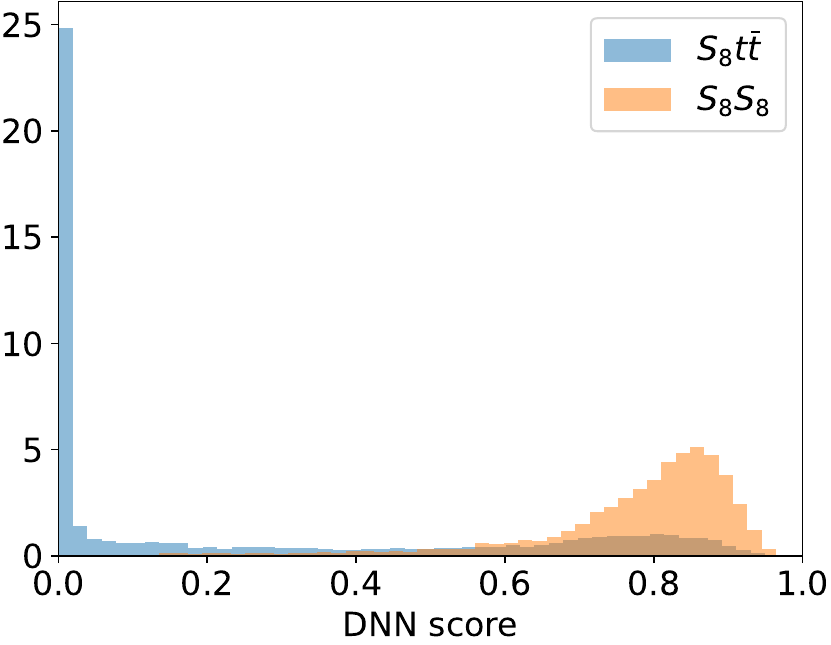}
    \end{subfigure}
    \begin{subfigure}{0.32\linewidth}
        \includegraphics[width=\linewidth]{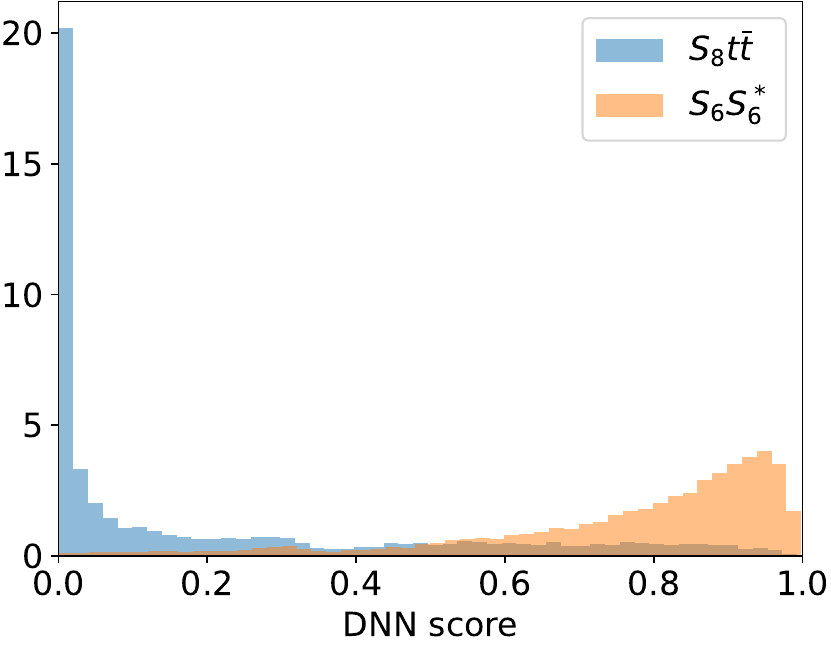}
    \end{subfigure}

    \begin{subfigure}{0.32\linewidth}
        \includegraphics[width=\linewidth]{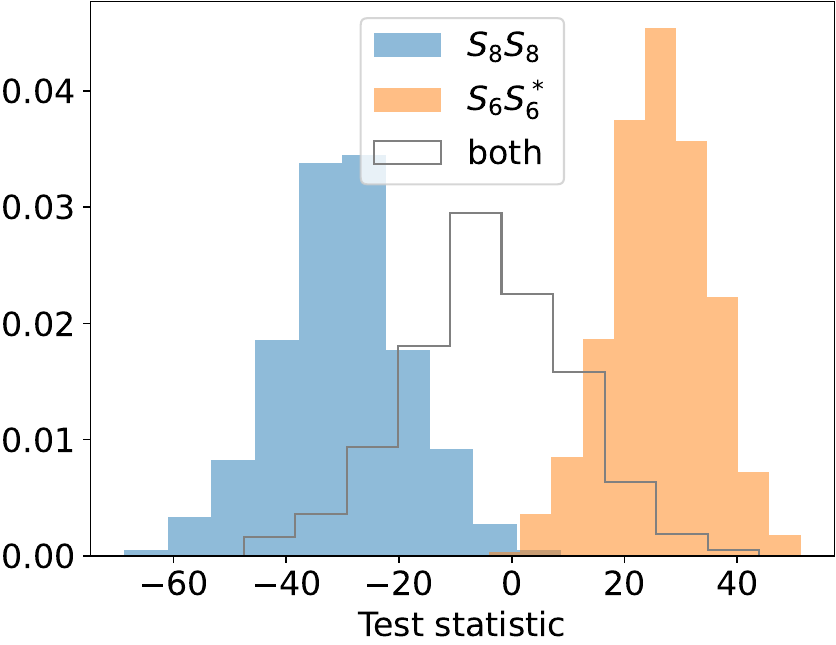}
        \caption{$S_8 S_8$ vs $S_6 S_6^*$}
        \label{fig:separation_a}
    \end{subfigure}
    \begin{subfigure}{0.32\linewidth}
        \includegraphics[width=\linewidth]{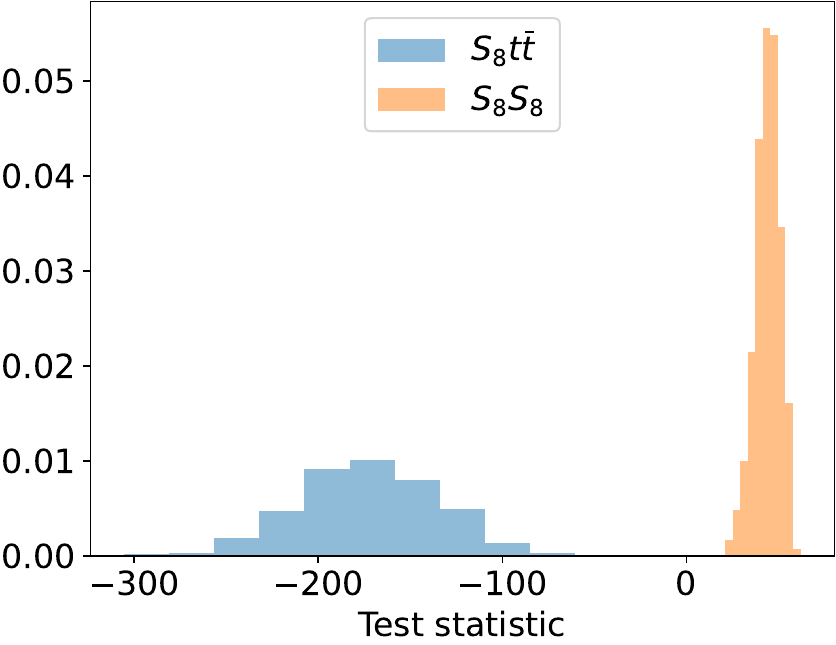}
        \caption{$S_8 t\bar t$ vs $S_8 S_8$}
        \label{fig:separation_b}
    \end{subfigure}
    \begin{subfigure}{0.32\linewidth}
        \includegraphics[width=\linewidth]{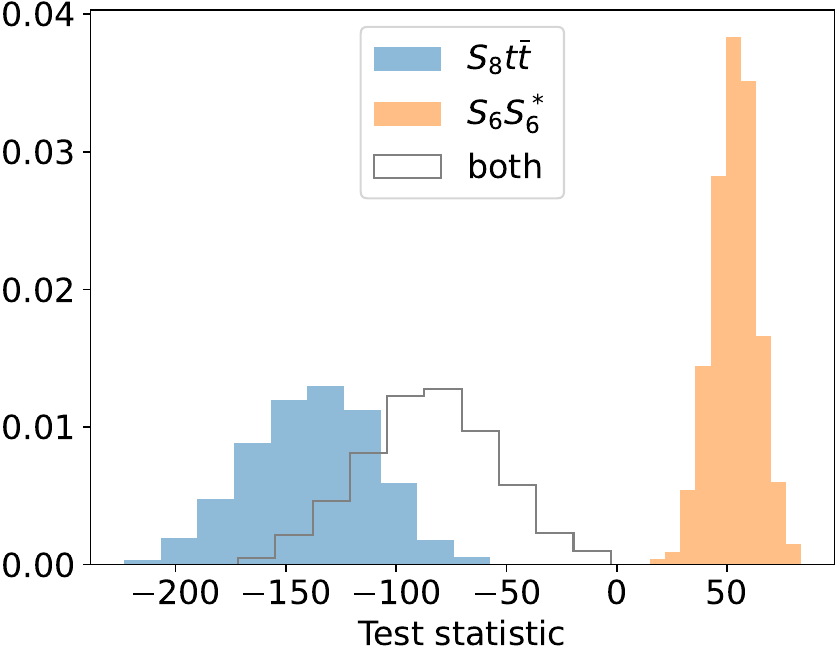}
        \caption{$S_8 t\bar t$ vs $S_6 S_6^*$}
        \label{fig:separation_c}
    \end{subfigure}
    \caption{Separating two signal processes for a scalar mass of 1.8~TeV. The top row shows the NN score distribution obtained from training the CNN to separate two signal processes. The second row shows the distributions of the corresponding test statistics as defined in \cref{eq:teststatistic}. }
    \label{fig:separation}
\end{figure}

As a first step we use the results of the signal vs.\ background discrimination described in the previous section to isolate for each process the signal events that the network identifies with a high confidence. 
Specifically, we place a cut on the NN score such that the remaining signal events have at least a significance of $5\sigma$.
Next, we train our CNN architecture to separate each pair of signal processes, resulting in the normalised NN score distributions in the top row of \cref{fig:separation}.
We take these distributions as probability density functions and refer to the PDF of the process $j$ as $f_j(x)$.
We then draw samples $x_i$ from this distribution where $i=1,\ldots , N_{5\sigma}$ with $N_{5\sigma}$ the number of events needed for discovery.
The goal is to determine whether they originate from distribution $j=1$ or 2.
To this end we construct the test statistic $t$ from a log-likelihood ratio,
\begin{align}
    t = -2 \ln \frac{\mathcal L_1(\{x_i\})}{\mathcal L_2(\{x_i\})}, \qquad \mathcal L_j(\{x_i\}) = \prod_{i=1}^{N_{5\sigma}^j} f_j(x_i)\label{eq:teststatistic}
\end{align}
for two hypotheses $f_1$ and $f_2$.
By drawing many such sample sets and calculating the test statistic for each, we obtain the distribution of $t$, shown in the bottom row in \cref{fig:separation}.

The top row of \cref{fig:separation} already gives a visual indicator that each combination of signal events can be separated by the networks.
However, for a quantitative statement it is necessary to introduce the test statistic.
Whereas an excess consists of several events with DNN scores, we can calculate a single test statistic from it, which can then be compared to the distributions in the second row of \cref{fig:separation}.
From \cref{fig:separation_b,fig:separation_c} we can see that $S_8 t\bar{t}$ can be very clearly distinguished from pure pair production, and also the two pair production distributions in \cref{fig:separation_a} do not overlap.
An excess would therefore most likely fall within either distribution, thus identifying the signal process.
We also consider the case that both signal processes are present simultaneously -- a scenario that occurs naturally in composite Higgs models \cite{Cacciapaglia:2015eqa}.
This leads to the distributions outlined in gray, which lie between the pure signal hypotheses. 

\section{Conclusions and outlook}
\label{sec:conclusion}
In this article we studied prospects to search for pair or singly produced colour octet or colour sextet scalars which decay dominantly into two top quarks at the LHC. We focused on the same-sign lepton final state. After applying common preselection criteria on Monte Carlo simulated signal and background events, we trained a neural network comprising of a simple multilayer perceptron  combined with a convolutional neural network to optimize the separation of signal and background events. For LHC operated at 14 TeV and a luminosity of 3 ab$^{-1}$ we find (see \cref{fig:results}) an expected discovery reach of $m_8=1.8$~TeV and $m_6=1.92$~TeV for pair produced colour octets and sextets and an expected exclusion reach of $m_8=2.02$~TeV and $m_6=2.14$~TeV. For a sizable coupling of the colour resonance to two tops, the $4t$ final state production cross section gets further enhanced due to production of a single colour resonance associated with two tops which increases the discovery reach to even higher masses.

Retraining the same network architecture on model discrimination, we showed for a benchmark point of $m=1.8$~TeV that the signal events required for discovery are sufficient to clearly discriminate the different signal candidates: colour octet pair production, colour octet single production and colour sextet pair production (see \cref{fig:separation}).

The methodology and networks do not depend on the precise nature of the produced states but only on the final state $tt\bar{t}\bar{t}$. Thus, they can be applied to different spin and colour structures, e.g.~to the production of colour neutral scalar or spin-1 states as well as colour spin-1 states\footnote{These states are commonly present for example in composite Higgs models with an underlying fermionic description. See \cite{Cacciapaglia:2024wdn} for a classification.}.

\section*{Acknowledgements}
We thank Raimund Ströhmer for useful discussions about separating the signal processes and Benjamin Fuks for discussions on Ref.~\cite{Darme:2024epi}. This work is supported by the Center for Advanced Computation at Korea Institute for Advanced Study. T.F.~is supported by a KIAS Individual Grant (AP083701) via the Center for AI and Natural Sciences at the Korea Institute for Advanced Study. 
M.K.~has been supported by the ``Studienstiftung des deutschen Volkes'' and the DFG research training group GRK2994. 
W.P.~has been supported by DFG, project no.~PO-1337/12-1.
JHK and JP are supported partly by the National Research Foundation of Korea (NRF) Grant NRF-2021R1C1C1005076, the BK-21 FOUR program through NRF, and the Institute of Information \& Communications Technology Planning \& Evaluation (IITP)-Information Technology Research Center (ITRC) Grant IITP-2025-RS-2024-00437284. JP also receives support from the NRF, under grant number RS-2024-00394030.


\clearpage
\appendix

\clearpage

\section{Details on the event generation}\label{app:single}

\begin{figure}
    \centering
    \begin{subfigure}{0.47\linewidth}
        \includegraphics[width=\linewidth]{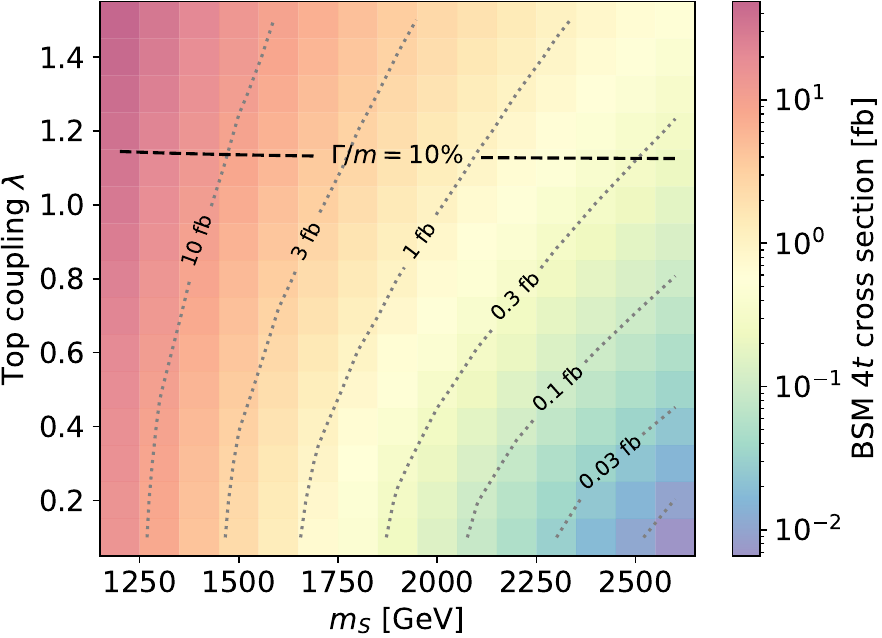}    
        \caption{$S_8$ single + pair production}
        \label{fig:xssinglepair}
    \end{subfigure}\quad
    \begin{subfigure}{0.47\linewidth}
        \includegraphics[width=\linewidth]{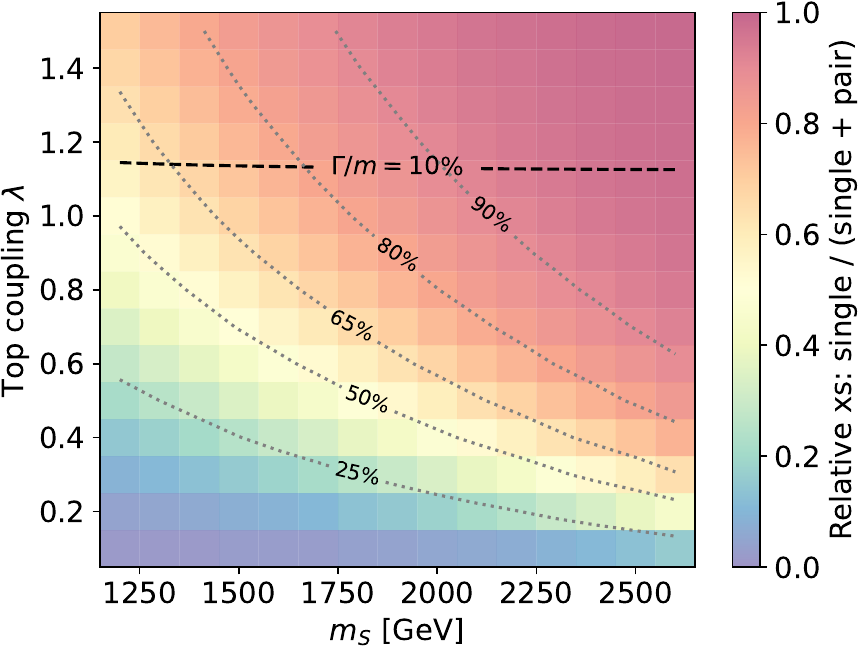}
        \caption{Relative contribution of single production}
        \label{fig:singlerelativexs}
    \end{subfigure}
    \caption{Cross sections for producing four top quarks via coloured scalars.
    For the octet we show (a) the absolute cross section and (b) the percentage of the cross section that is due to single production.
    The horizontal dashed line indicates the validity region of the narrow width approximation.}\label{fig:cross_sections}
\end{figure}

In the main text we studied the colour octet scalars with a decay to two top quarks.
This coupling to top quarks allows for the associated production $pp\to t\bar t S_8$ as an additional production channel in addition to QCD pair production.
Note, however, that splitting the production cross section into a single and a pair production contribution is -- strictly speaking -- unphysical. The QCD pair production contains a $gg$ initial state contribution, i.e.\ it has the same initial and final state as the single production contribution and is of the same order in all couplings. Thus, there is no \emph{physical} distinction between single production and $gg$ initial state pair production (in particular if one of the $S_8$ is off-shell). Also, on the technical level, we cannot generate events for only single production within the \textsc{MadGraph5} framework as the command corresponding to the single production diagram in \cref{fig:feynman_4t_v2} always generates pair production events as well\footnote{For technical details see \url{https://answers.launchpad.net/mg5amcnlo/+question/689616} and \url{https://bugs.launchpad.net/mg5amcnlo/+bug/1367374}.}.
We thus have to clarify what we refer to as ``single production'' and ``pair production'' cross sections:
\begin{itemize}
\item{For ``pair production'' we generate datasets with $\lambda_8\lll 
 1$. We keep the coupling just large enough to avoid displaced decays of $S_8\rightarrow t\bar{t}$. In this case, the contribution to the cross section from the single production diagram in \cref{fig:feynman_4t_v2} as well as from off-shell $S_8$ pair production are negligible.}
 \item{For ``single production'' we generate datasets with large $\lambda_8 = 1.1$.}
 \end{itemize}
In \cref{fig:xssinglepair} we show the dependence of the cross section on mass and coupling. As can be seen, for $\lambda_8 = 1.1$, the width of $S_8$ is still sufficiently narrow to work in the narrow width approximation (NWA). In \cref{fig:singlerelativexs} we show the relative contribution of single and pair production to the cross section. As a proxy for the single production cross section we use $\sigma(pp\to 4t; S_8)- \sigma(pp\to S_8 S_8)$, where ``$;S_8$'' indicates that at least one $S_8$ has to occur. This is only a good approximation if interference effects are small. For $\lambda_8 = 1.1$, single production clearly dominates over pair production in the $m_S$ regime of interest for this article.

\section{Details on the separation of signal processes}\label{app:sep}

\begin{figure}
    \centering
    \begin{subfigure}{0.31\linewidth}
        \includegraphics[width=\linewidth]{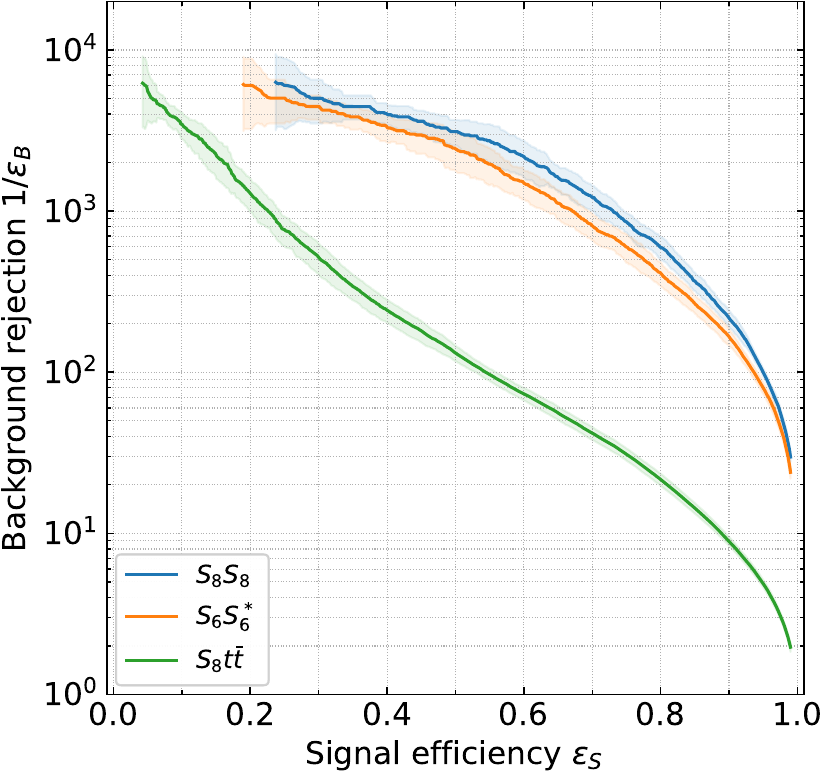}
        \caption{$S_8S_8$}
        \label{fig:roc2_a}
    \end{subfigure}
    \begin{subfigure}{0.31\linewidth}
        \includegraphics[width=\linewidth]{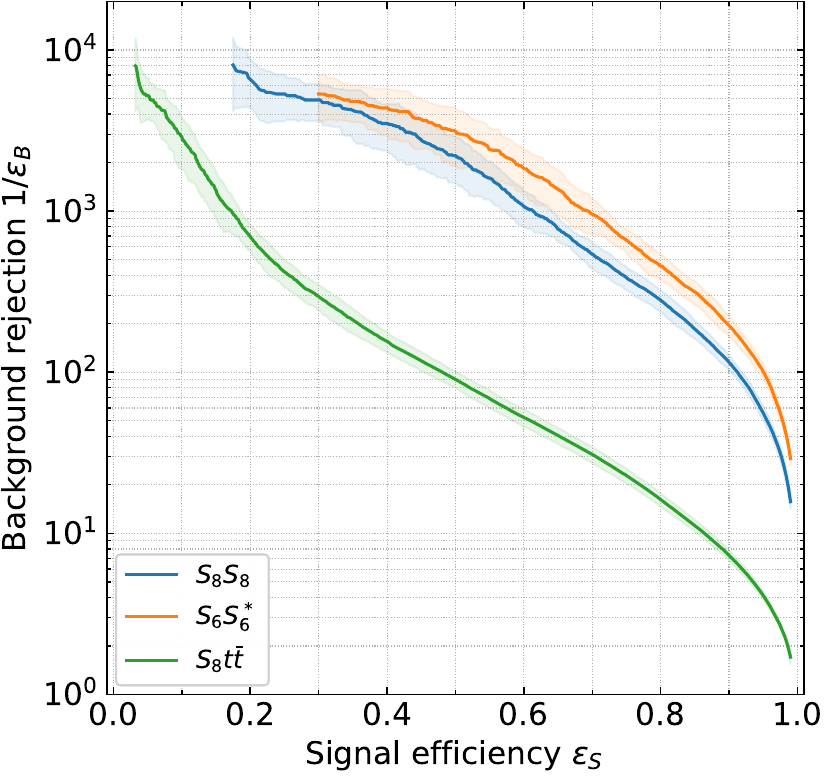}
        \caption{$S_6S_6^*$}
        \label{fig:roc2_b}
    \end{subfigure}
    \begin{subfigure}{0.31\linewidth}
        \includegraphics[width=\linewidth]{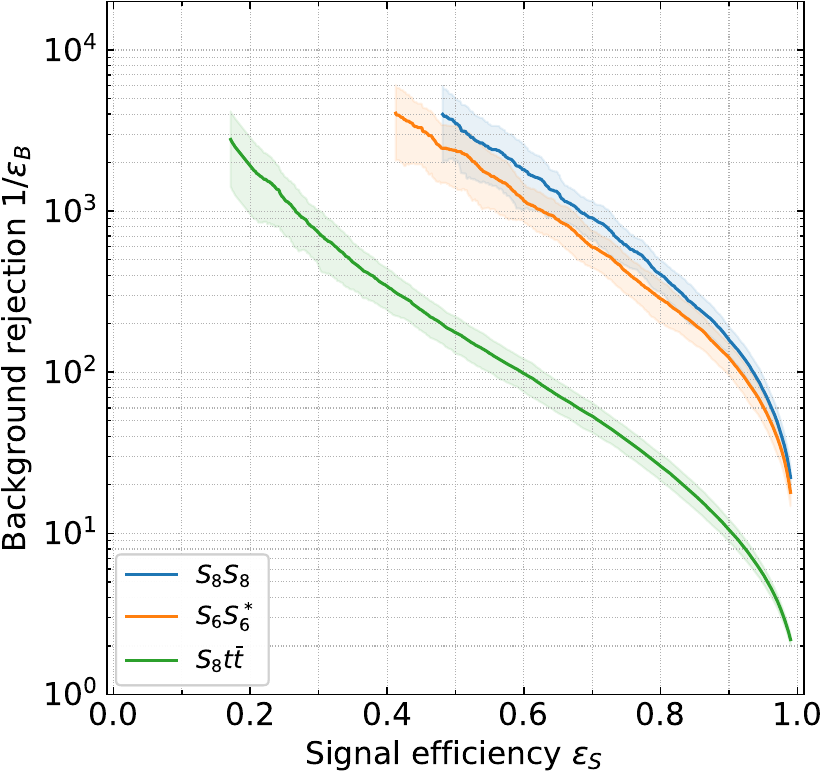}
        \caption{$S_8t\bar{t}$}
        \label{fig:roc2_c}
    \end{subfigure}
    \caption{Receiver operator characteristic (ROC) curves comparing networks trained on  $S_8S_8$, $S_6S_6^*$, $S_8t\bar{t}$, tested across different signal types.}
    \label{fig:roc2}
\end{figure}

In \cref{sec:seperation} we claimed that there is substantial cross acceptance among $S_8 S_8$, $S_6 S_6^*$, and $S_8 t\bar{t}$ signal events when training networks on separating these signal events from SM background events. In this appendix, we quantify this statement. \cref{fig:roc2} shows the ROC curves resulting from training to separate SM backgrounds from  (\cref{fig:roc2_a}) $S_8S_8$ signal, (\cref{fig:roc2_b}) $S_6S^*_6$ signal, and (\cref{fig:roc2_c}) $S_8t\bar{t}$ signal for a benchmark mass $m_S=1.8$~TeV. For each case, we show the ROC curves obtained when applying the trained networks to $S_8 S_8$, $S_6 S_6^*$, and $S_8 t\bar{t}$ test datasets. As can be seen from \cref{fig:roc2_a} the network trained on $S_8 S_8$ has highest signal efficiency on $S_8 S_8$ test data. However, the network shows almost as high signal efficiency for $S_6 S^*_6$ test data while $S_8 t\bar{t}$ signal efficiency is substantially lower. \cref{fig:roc2_a} show analogous results with reversed r\^oles of $S_8 S_8$ and $S_6 S^*_6$. This behaviour is to be expected as the networks are trained on separation from background for which e.g. $S_T$ and the brightness of jet images (corresponding to high $p_T$) play a dominant r\^ole, while angular correlations between the same-sign leptons (which is an example for an important discriminator between $S_8 S_8$ and $S_6 S^*_6$ signals, see \cref{fig:jetimages}) are less relevant. This observation motivates the re-training for model discrimination as performed in \cref{sec:seperation}. The ROC curves in \cref{fig:roc2_c} from the network trained on $S_8t\bar{t}$ might appear surprising as efficiencies for pair production test data sets exceed those of the $S_8t\bar{t}$ test data set although the network was trained on single production. The result reflects simply that single production events are harder to distinguish from SM backgrounds (one reason is the lower $S_T$ and lower $p_T$ of jets and leptons resulting from the decay of associated tops as compared the tops resulting from pair production of heavy coloured scalars and their decay in tops) even if the training is focussed on them.  Note, that the $S_8t\bar{t}$ efficiency in \cref{fig:roc2_c} in exceeds those in \cref{fig:roc2_a} and \cref{fig:roc2_b} which shows that the network is better optimized for single production than the pair production trained networks.

\section{Neural network architectures}
\label{sec:appen}

In this section, we outline the various network architectures we explored and assess their performance through an analysis of the corresponding ROC curves. 

\subsection{Simple multilayer perceptron}
\label{sec:MLP}
\begin{figure*}[t]
	\centering
	\begin{center}
		\includegraphics[width=0.5\textwidth,clip]{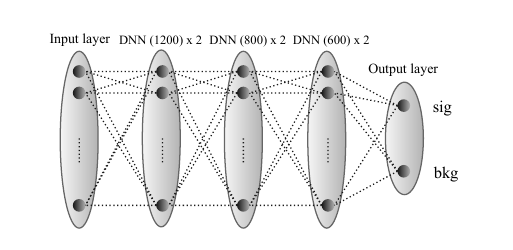}
		\caption{A schematic architecture of the simple multilayer perceptron (MLP) used in this article.}
		\label{fig:MLP} 
	\end{center}
\end{figure*}

MLPs are simple structures that can easily be used when the input consists of the four-momenta of reconstructed particles or combinations of kinematic variables. As illustrated in \cref{fig:MLP}, each neuron in a layer is fully connected to every neuron in the subsequent layer. The implemented MLP consists of 6 hidden layers with a gradually decreasing number of neurons from 1200 to 300, followed by an output layer.
The output layer contains two neurons, each representing a different class, such as a background and a signal. The values of these two neurons are denoted by $\hat{p}_k$, where $k=0$ or 1 represents the respective classes.
We normalise the values $\hat{p}_k$ as $p_{k} = e^{\hat{p}_{k}}/\sum e^{\hat{p}_{k}}$ using a softmax function, where $p_0$ represents the probability that the network assigns the event to class 0, and similarly for $p_1$ with class 1.
We train the network using the cross entropy loss function,
\begin{eqnarray}
L = -y_k \log p_k - (1-y_k) \log (1-p_k) \;,
\end{eqnarray} 
where events of class 0 and class 1 are labeled as $y_0 = 0$ and $y_1 = 1$, respectively. 
Training is performed with the {\sc Adam} optimizer with a mini-batch size of 20 and a learning rate of $1 \times 10^{-4}$. To avoid overfitting, we monitor the validation loss and train the model for up to 1000 epochs. If the validation loss does not improve within 30 epochs, training is stopped, and the parameters corresponding to the epoch with the minimal validation loss are saved\footnote{Unless otherwise stated, the activation function, the type of optimiser, the loss function, the configuration of an output layer, and the use of a validation set are the same for all other neural networks.}.

\subsection{Convolutional neural networks}
\label{sec:CNN}
\begin{figure*}[t]
	\centering
	\begin{center}
		\includegraphics[width=1.\textwidth,clip]{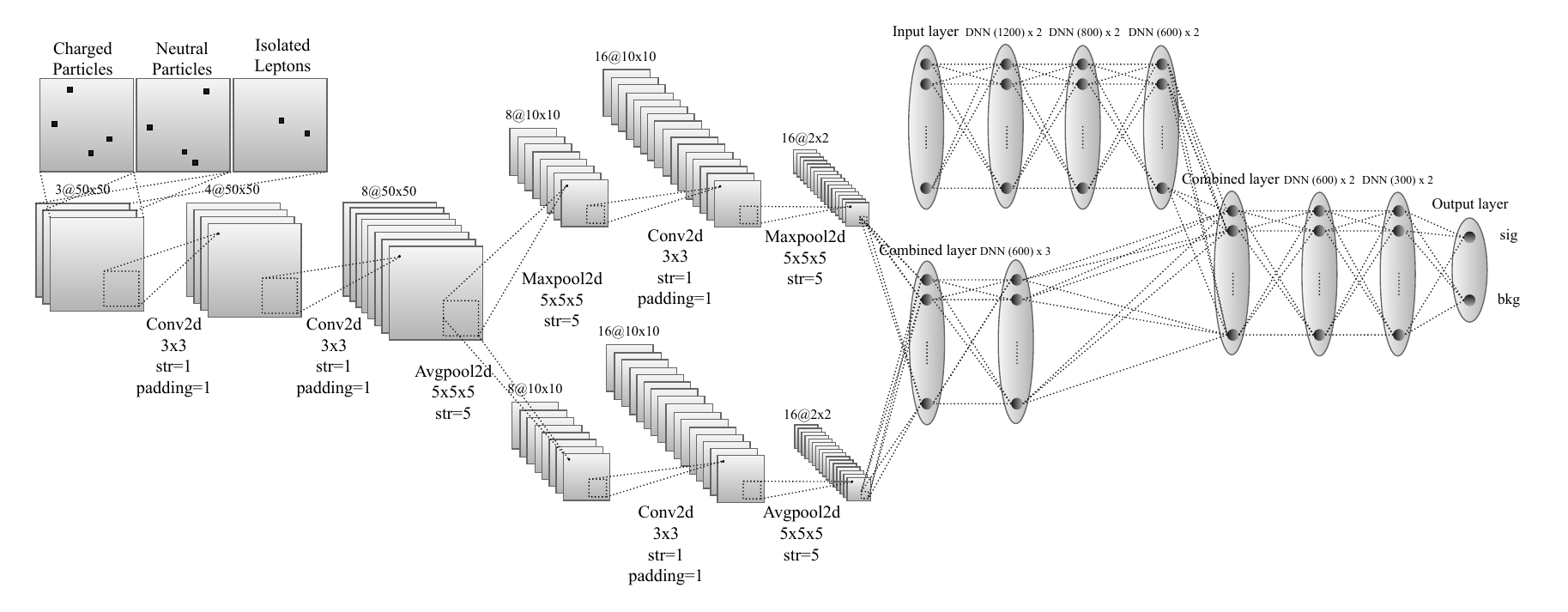}
		\caption{A schematic CNN architecture  used in this article. The separate MLP in the right-upper corner is used on the kinematic dataset $\mathcal{K}$.}
		\label{fig:CNN} 
	\end{center}
\end{figure*}

CNNs are well-suited for image recognition tasks, especially when the final state is represented as images. The input is a 3D image of $\mathcal I_{CN\ell}$ with dimensions $ 3 \times 50 \times 50 $ where 3 indicates the number of image layers as specified in \cref{fig:CNN}.

We first apply two consecutive 2D convolutions, each with kernel size $ 3 \times 3 $, stride of 1, padding of 1. The number of feature maps increases from 4 in the first convolution to 8 in the second. Each convolution is followed by batch normalization and the Rectified Linear Unit (ReLU) \cite{pmlr-v15-glorot11a}  activation function. The resulting image size has dimensions $ 8 \times 50 \times 50$.

In CNNs, the pooling operation selects the most important pixel in each region of a feature map, such as using max pooling to retain the maximum value or average pooling for the mean value, while discarding the rest. This allows CNNs to reduce the input image dimensions and extract key features at different scales through combined convolution and pooling operations.

As illustrated in \cref{fig:CNN}, we implement a two-branch architecture that utilizes both max and average pooling, each with a kernel size of $5 \times 5$ and stride of 5, reducing the image size to $8 \times 10 \times 10$. Each branch then undergoes another 2D convolution and pooling, ultimately reducing the dimensions to $16 \times 2 \times 2$. The final output is flattened to a one-dimensional vector and connected to 3 hidden layers, each containing 600 neutrons.

The above CNN is combined with a MLP that takes kinematic variables as inputs as detailed in the previous subsection. The MLP is merged with the last hidden layer of the CNN, followed by 4 additional hidden layers with neurons decreasing from 600 to 300. The output layer is connected to the last hidden layer. Training is performed with a mini-batch size of 20 and a learning rate of $1 \times 10^{-4}$.


\bibliographystyle{JHEP}
\bibliography{bibliography}

\end{document}